\pdfoutput=1

\documentclass[11pt]{article}

\usepackage[final]{acl}

\usepackage{times}
\usepackage{latexsym}

\usepackage[T1]{fontenc}

\usepackage[utf8]{inputenc}

\usepackage{microtype}

\usepackage{amsmath, amscd, amssymb, amsthm}
\usepackage{array}
\usepackage{graphicx}
\usepackage{bm}
\usepackage{bbm}
\usepackage{booktabs}
\usepackage{enumitem}
\usepackage{footnote}
\usepackage{mathtools}
\usepackage{multicol}
\usepackage{multirow}
\usepackage{natbib}
\usepackage{outlines}
\usepackage{pifont}%
\usepackage{subcaption}
\usepackage{soul}
\usepackage{tabularx}
\usepackage{tabulary}
\usepackage{url}
\usepackage{xcolor}
\usepackage[utf8]{inputenc}
\usepackage{hyperref}
\usepackage{cleveref}
\usepackage{diagbox}
\usepackage{tablefootnote}

\usepackage{header}

\newcommand{\name}{GTR}
\newcolumntype{R}{>{\raggedleft\arraybackslash}p{3em}}
\newcolumntype{C}{>{\centering\arraybackslash}p{3em}}

\newcommand{\cmark}{\ding{51}}%
\newcommand{\xmark}{\ding{55}}%

\title{Large Dual Encoders Are Generalizable Retrievers}

\author{Jianmo Ni, Chen Qu, Jing Lu, Zhuyun Dai, \\ \textbf{Gustavo Hern\'{a}ndez \'{A}brego, Ji Ma, Vincent Y. Zhao, Yi Luan,} \\ \textbf{Keith B. Hall, Ming-Wei Chang, Yinfei Yang} \\
{\rm Google Research}\\Mountain View, CA
}

\begin{document}
\maketitle
\begin{abstract}
It has been shown that dual encoders trained on one domain often fail to generalize to other domains for retrieval tasks. One widespread belief is that the bottleneck layer of a dual encoder, where the final score is simply a dot-product between a query vector and a passage vector, is too limited to make dual encoders an effective retrieval model for out-of-domain generalization. 
In this paper, we challenge this belief by scaling up the size of the dual encoder model {\em while keeping the bottleneck embedding size fixed.} With multi-stage training, surprisingly, scaling up the model size brings significant improvement on a variety of retrieval tasks, especially for out-of-domain generalization. Experimental results show that our dual encoders, \textbf{G}eneralizable \textbf{T}5-based dense \textbf{R}etrievers (GTR), outperform %
existing sparse and dense retrievers on the BEIR dataset~\cite{thakur2021beir} significantly. Most surprisingly, our ablation study finds that GTR is very data efficient, as it only needs 10\% of MS Marco supervised data to achieve the best out-of-domain performance.\footnote{All the GTR models are released at \url{https://tfhub.dev/google/collections/gtr/1}.}

\end{abstract}

\section{Introduction}

Typical neural retrieval models follow a dual encoder paradigm~\cite{Gillick2018EndtoEndRI,yang-etal-2020-multilingual,karpukhin-etal-2020-dense}.
In this setup, queries and documents are encoded separately into a shared fixed-dimensional embedding space where relevant queries and documents are represented in each other's proximity. Then, approximated nearest neighbor search~\cite{ann1,ann2} is applied to efficiently retrieve relevant documents given an encoded input query.

\begin{figure}[t]
  \centering
  \includegraphics[width=\linewidth]{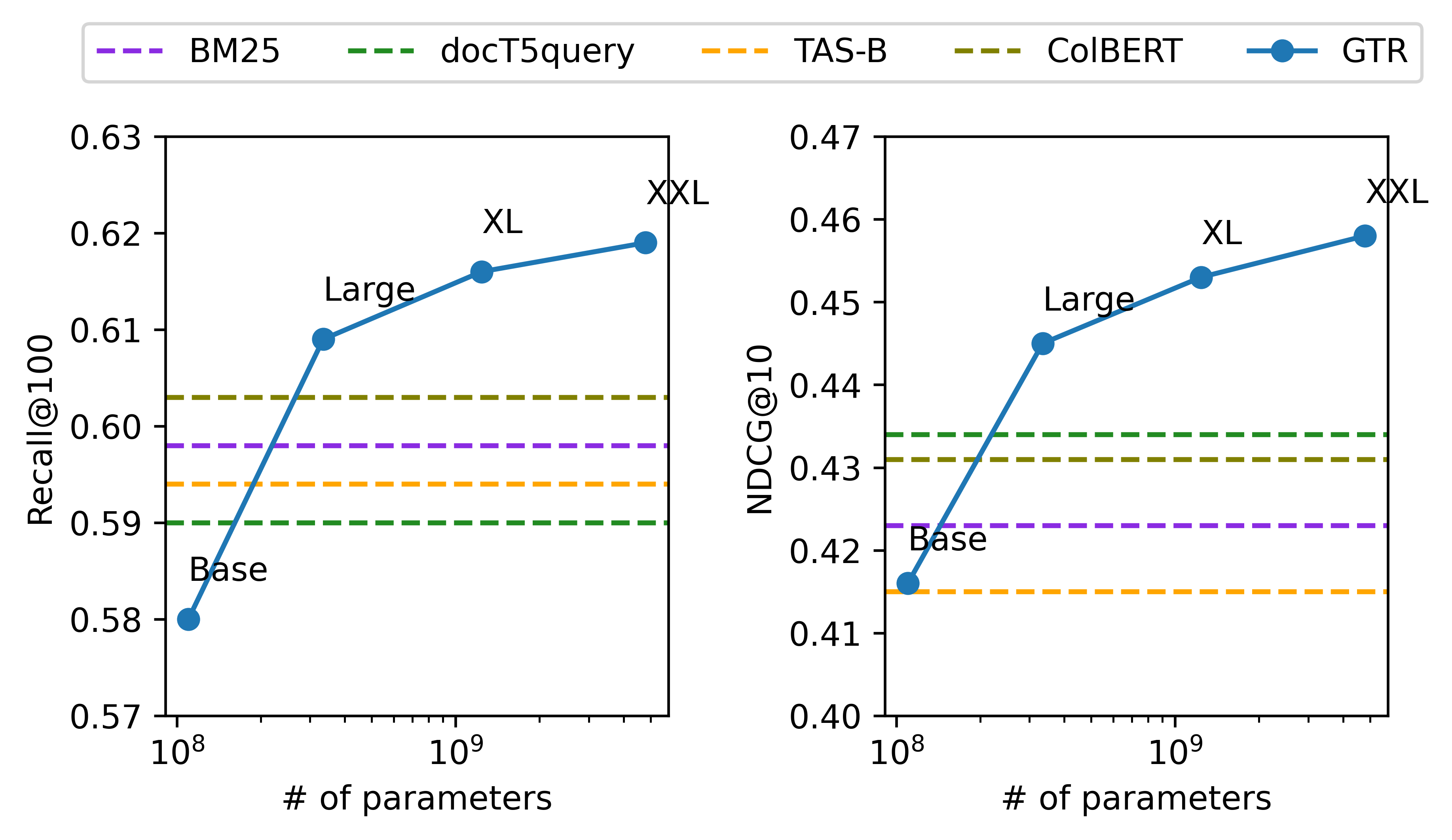}
  \caption{The average Recall@100 and NDCG@100 on all BEIR tasks excluding MS Marco. Scaling up consistently improves dual encoders' out-of-domain performance.
  }
  \label{fig:gtr_avg}
\end{figure}

\begin{figure*}
  \centering
  \includegraphics[width=0.9\linewidth]{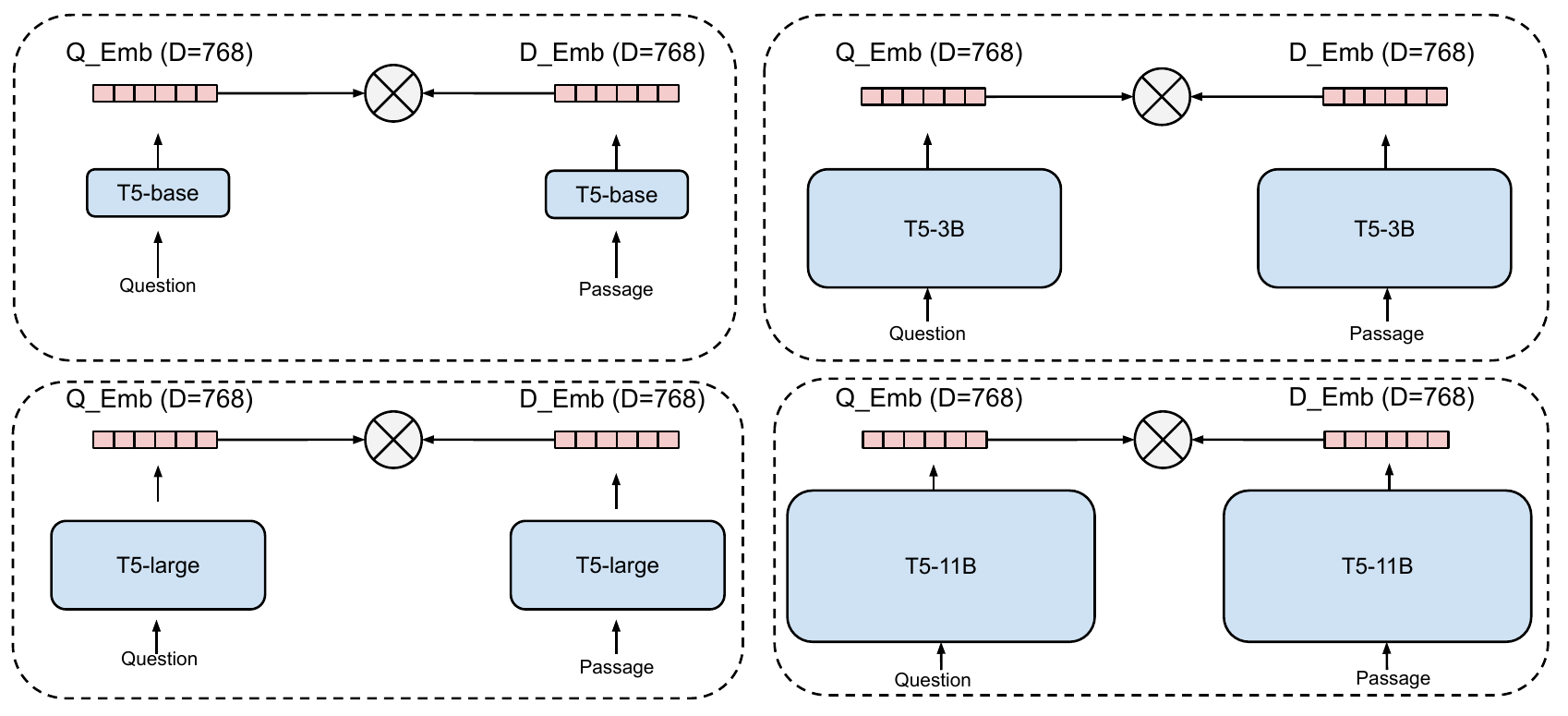}
  \caption{Architecture of {\bf G}eneralizable {\bf T}5-based dense {\bf R}etrievers. The research question we ask is: {\em can scaling up dual encoder model size improve the retrieval performance while keeping the bottleneck layers  {\bf \em fixed?}} Only the encoder is taken from the pre-train T5 models, and the question tower and document tower of the dual encoder share parameters.}
  \label{fig:de}
\end{figure*}

While dual encoders are popular neural retrievers,
the expressiveness of the model is limited by
a bottleneck layer consisting of only a simple dot-product between query embeddings and passage embeddings.
Several papers~\cite{lu2021multi, khattab2020colbert} have discussed that the simple dot-product (or cosine similarity) between the embeddings might not be powerful enough to capture semantic relevance. \citet{thakur2021beir} studied whether the retriever models can generalize to other domains and conclude that
dual encoder models have ``issues for out-of-distribution data'', and showed that models with more interactions between queries and documents have better generalization ability.

In this paper, we challenge this belief by scaling up the dual encoder model size while keeping the bottleneck embedding size fixed. Note that scaling up a dual encoder is different from scaling up pretrained language models such as BERT~\cite{Devlin2019BERTPO} and T5~\cite{2020t5} because of the presence of the bottleneck layer. While increasing the model size can greatly increase the capacity of the model, for dual encoders, where the embedding size is fixed, the interactions between queries and documents are still limited by a simple dot-product.

In order to test this hypothesis, we take advantage of the existing T5 model architecture and checkpoints, which allows us to build encoders of up to 5 billion parameters while keeping the bottleneck embedding dimension of 768 in all configurations, as illustrated in Figure~\ref{fig:de}. Following~\citet{ni2021sentence}, we build dual encoders by taking the encoder part of T5. For effectively using the power of large models, we collect roughly two billion community question-answer pairs as generic pre-training data. 
By combining pre-training using generic training data and fine-tuning using MS Marco~\cite{nguyen2016msmarco}, we are able to train large-scale dual encoder retrieval models.
We call the resulting models \textbf{G}eneralizable \textbf{T}5-based dense \textbf{R}etrievers~(\textbf{\name}).

We evaluate the zero-shot performance of GTR on the BEIR benchmark~\cite{thakur2021beir}, which consists of 18 selected information retrieval tasks across 9 domains.\footnote{We focus on evaluating the performance on the 18 BEIR~\cite{thakur2021beir} tasks other than MS Marco and we did not use in-domain training data or question generation~\cite{ma-etal-2021-zero}.} Our results show that, surprisingly, 
scaling up of dual encoders leads to better generalizability despite the fixed bottleneck embedding dimension.
Second, pre-training on community question-answer pairs and fine-tuning on human curated data are both important to fully utilize the power of the scaled up model. 
In addition, with scaling and pre-training, we found GTR to be highly data efficient in terms of human annotated queries, as it only needs to use 10\% of MS Marco to match the overall out-of-domain performance.

\section{Background}
\subsection{Dual Encoder and dense retrieval}

Classic retrieval models such as BM25~\cite{bm25} relies on lexical overlap, term frequency heuristics, inverse document frequency and document length. This type of retrieval models does not require any training and can generalize reasonably well due to its emphasis on lexical match. However, these retrieval models fall short of finding documents that are only semantically related to the query and have low lexical overlap. %

This is where the dense retrieval models come into play. The retrieval process is \textit{dense} because both queries and documents are embedded into low-dimensional dense representations, in contrast to the high-dimensional sparse representations used in lexical based retrieval functions. Such an encoding process is often accomplished by a dual encoder architecture, with one of the encoders tailored to the queries and the other to the documents. Matching using dense representations enables dense retrieval models to go beyond lexical overlap to retrieve semantically similar documents. This powerful capability of semantic matching, however, often requires relatively large amounts of training data.

Another critical challenge for dual encoder models is that the performance is possibly bounded by the dot-product similarity function. As such, there is growing interest in applying lightweight interaction layers to replace the single dot-product function. On the one hand, \newcite{luan2020sparse} proposes a multi-vector encoding model, which represents each document as a fixed-size
set of multiple vectors, and calculate the relevance scores as the maximum inner product over this set. This model combines the efficiency of dual encoders with some of the expressiveness of attention based architectures. On the other hand, ColBERT \citep{khattab2020colbert} proposes to learn embeddings for each token and then use a ``MaxSim'' operation to select the best candidate. These models achieve significant improvement but also introduce a large latency overhead. In this paper, we take a step back and aim to empirically study how to improve the performance of single dot-product based methods.  Specifically, we study whether simply scaling up the model capacity can lead to better fixed embeddings to improve the performance of single dot-product retrievers.

\subsection{BEIR generalization task}
\label{sec:beir}

For evaluation in this paper we use BEIR, a heterogenous benchmark for zero-shot evaluation of information retrieval models.
The BEIR zero-short evaluation suit contains 18 information retrieval datasets\footnote{MS Marco is excluded from the zero-shot comparison as many baseline model used it as training data.} 
across 9 domains, including \textit{Bio-Medical}, \textit{Finance}, \textit{News}, \textit{Twitter}, \textit{Wikipedia}, \textit{StackExchange}, \textit{Quora}, \textit{Scientific}, and \textit{Misc}.
The majority of the datasets have binary relevancy labels indicating whether a document is relevant to a given query or not. A small part of the datasets have 3-level or 5-level relevancy judgements. 
We refer readers to the original BEIR paper~\cite{thakur2021beir} for more details.

\section{Generalizable T5 Retriever}

\subsection{T5 dual encoder}

We use the dual encoder framework to train dense retrieval models.  
We follow prior work \citep{xiong2020approximate,hofstatter2021efficiently} to initialize dual encoders from pre-trained language models. 
In this work, we found convenient to use the pre-trained T5 model family as our backbone encoder because the T5 model family provides off-the-shelf pre-trained models (e.g.~T5, mT5, byT5) with a wide range of model capacity from millions to billions of parameters \citep{2020t5,xue2020mt5,xue2021byt5}. 
The architectures of our models are illustrated in \cref{fig:de}. 

Let paired examples $\mathcal{D}=\{(q_i, p_i^{+})\}$ be the training set, where $q_i$ is an input question and $p_i^{+}$ is a related passage (e.g., a semantically relevant passage to the question). 
Following \citet{ni2021sentence}, we encode the question $q_i$ and passage $p_i^{+}$ into embeddings by feeding them to the T5 encoder and taking the mean pooling of the encoder as output. In all outr experiments, we fix the output embeddings to be of size 768.

We train the model using an in-batch sampled softmax loss~\citep{Henderson2017EfficientNL}:
\begin{equation}
  \mathcal{L} = \frac{e^{\text{sim}(q_i, p_i^{+})/ \tau}}{\sum_{j \in \mathcal{B}} { e^{\text{sim}(q_i, p_j^{+}) / \tau} } },
  \label{eq::loss}
\end{equation}
where the similarity scoring function \textit{sim} is the cosine similarity between the embeddings of $q_i$ and $p_i^{+}$.
$\mathcal{B}$ is a mini-batch of examples and $\tau$ is the softmax temperature. 

Additional negatives $p_j^{-}$ can be given for input question $q$.  The loss is computed by including them in the denominator:
\begin{equation}
  \mathcal{L} = \frac{e^{\text{sim}(q_i, p_i^{+})/ \tau }}{\sum_{j \in \mathcal{B}} { e^{\text{sim}(q_i, p_j^{+})/ \tau}  + e^{\text{sim}(q_i, p_j^{-})/ \tau} }} .
\end{equation}

We also apply a bi-directional in-batch sampled softmax loss, where we compute losses for both question to document matching and document to question matching.

\subsection{Multi-stage training}

\begin{figure}
    \centering
    \includegraphics[width=0.9\linewidth]{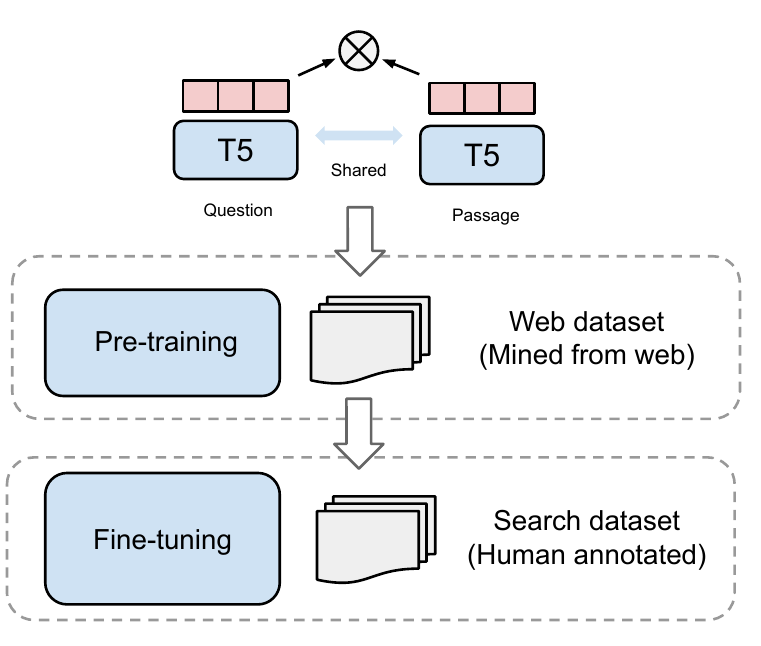}
    \caption{Multi-stage training for generalizable T5 dual encoder.}
    \label{fig:multi_stage}
\end{figure}

As shown in \cref{fig:multi_stage}, we use a multi-stage dual encoder training approach to achieve generalizable retrieval models. 

The training process includes a pre-training stage on a web-mined corpus and a fine-tuning stage on search datasets. 
The web-mined corpus provides a large amount of semi-structured data pairs (such as question-answer pairs and conversations), which can provide rich semantic relevance information. It is easy to collect but it is often not well annotated, if at all.
The search datasets are often annotated by humans, and the queries and documents are also authored by humans. These datasets are of high quality but costly to collect.

In this work, for dual encoder pre-training, we initialize the dual encoders from the T5 models and train on question-answer pairs collected from the Web. Recently, Sentence-T5 \citep{ni2021sentence}  explored different ways to extract strong text embeddings and achieved remarkable performance on SentEval and Sentence Textual Similarity tasks. We follow that setting to encode queries and passages via mean pooling from the T5 encoders and focus on the dense retrieval tasks.

For fine-tuning, our aim is to adapt the model to retrieval using a high quality search corpus so the model can learn to better match generic queries to documents. In this paper, we consider two datasets for fine-tuning: MS Marco \citep{nguyen2016msmarco} and Natural Questions \citep{kwiatkowski2019natural}.

\section{Experimental setup}

\subsection{Training Data}

\paragraph{Community QA.}
In order to leverage most of the power from the large scale models, we collect input-response pairs and question-answer pairs from online forums and QA websites including Reddit, Stack-Overflow, etc.
This results in 2 billion question-answer pairs that we use to pre-train the dual encoder.

\paragraph{MS Marco.} 
We consider the MS Marco dataset~\cite{nguyen2016msmarco}, which includes 532K query and document pairs, as search data for fine-tuning. 
The dataset is sampled from Bing search logs, which covers a broad range of domains and concepts. 
Most of the neural models compared in \cite{thakur2021beir} are trained on MS Marco, including DeepCT~\cite{deepct}, DocT5Query~\cite{Nogueira2019FromDT}, ANCE~\cite{xiong2020approximate} and ColBERT~\cite{khattab2020colbert}.
Some of these models have shown great generalization with comparable or even better performance relative to BM25.

\begin{table}[t]
\small
    \centering
    \begin{tabular}{|l|r|r|r|r|}
    GTR Models & Base & Large & XL & XXL \\
    \midrule
    \# of params & 110M & 335M & 1.24B & 4.8B \\
    \end{tabular}
    \caption{Number of parameters in the GTR models.}
    \label{tab:param}
\end{table}

\paragraph{Natural Questions.} %
In the fine-tuning stage, we also consider the Natural Questions dataset \citep{kwiatkowski2019natural} , which has been widely used in the dense retrieval literature \citep{karpukhin-etal-2020-dense,xiong2020approximate}. It consists of 130k query and passage pairs which are also human-annotated.

\begin{table}[t]
  \small
  \centering
  \resizebox{0.9\linewidth}{!}{%
\begin{tabular}{|l|r|}
Models & Dim. size  \\
\midrule
ColBERT & 128    \\
DPR, ANCE, TAS-B, GenQ, GTR  & 768   \\
BM25, DocT5Query & -  \\
  \end{tabular}
 }
  \caption{Dimension size of different models. Most dual encoder models set the embedding dimension size to 768.}
  \label{tab:dim_size}
\end{table}

\begin{table*}[t]
  \small
  \centering
  \resizebox{\linewidth}{!}{%
    \begin{tabular}{@{}l|cc|ccccc|cccc@{}}
\toprule
\multicolumn{1}{c|}{\multirow{2}{*}{NDCG@10 / Model}} & \multicolumn{2}{c|}{Lexical / Sparse}                & \multicolumn{5}{c|}{Dense}                                                                                                               & \multicolumn{4}{c}{Ours}                                                                                              \\ \cmidrule(l){2-12} 
\multicolumn{1}{c|}{}                                 & \multicolumn{1}{c|}{BM25}           & docT5query     & \multicolumn{1}{c|}{DPR}   & \multicolumn{1}{c|}{ANCE}  & \multicolumn{1}{c|}{TAS-B}       & \multicolumn{1}{c|}{GenQ}  & ColBERT        & \multicolumn{1}{c|}{GTR-Base} & \multicolumn{1}{c|}{GTR-Large} & \multicolumn{1}{c|}{GTR-XL}         & GTR-XXL        \\ \midrule
MS Marco                                              & \multicolumn{1}{c|}{0.228}          & 0.338          & \multicolumn{1}{c|}{0.177} & \multicolumn{1}{c|}{0.388} & \multicolumn{1}{c|}{0.408}       & \multicolumn{1}{c|}{0.408} & 0.401          & \multicolumn{1}{c|}{0.420}    & \multicolumn{1}{c|}{0.430}     & \multicolumn{1}{c|}{0.439}          & \textbf{0.442} \\
\midrule
Trec-Covid                                            & \multicolumn{1}{c|}{0.656}          & \textbf{0.713} & \multicolumn{1}{c|}{0.332} & \multicolumn{1}{c|}{0.654} & \multicolumn{1}{c|}{0.481}       & \multicolumn{1}{c|}{0.619} & 0.677          & \multicolumn{1}{c|}{0.539}    & \multicolumn{1}{c|}{0.557}     & \multicolumn{1}{c|}{0.584}          & 0.501          \\
BioASQ                                                & \multicolumn{1}{c|}{0.465}          & 0.431          & \multicolumn{1}{c|}{0.127} & \multicolumn{1}{c|}{0.306} & \multicolumn{1}{c|}{0.383}       & \multicolumn{1}{c|}{0.398} & \textbf{0.474} & \multicolumn{1}{c|}{0.271}    & \multicolumn{1}{c|}{0.320}     & \multicolumn{1}{c|}{0.317}          & 0.324          \\
NFCorpus                                              & \multicolumn{1}{c|}{0.325}          & 0.328          & \multicolumn{1}{c|}{0.189} & \multicolumn{1}{c|}{0.237} & \multicolumn{1}{c|}{0.319}       & \multicolumn{1}{c|}{0.319} & 0.305          & \multicolumn{1}{c|}{0.308}    & \multicolumn{1}{c|}{0.329}     & \multicolumn{1}{c|}{\textbf{0.343}} & 0.342          \\
NQ                                                    & \multicolumn{1}{c|}{0.329}          & 0.399          & \multicolumn{1}{c|}{0.474} & \multicolumn{1}{c|}{0.446} & \multicolumn{1}{c|}{0.463}       & \multicolumn{1}{c|}{0.358} & 0.524          & \multicolumn{1}{c|}{0.495}    & \multicolumn{1}{c|}{0.547}     & \multicolumn{1}{c|}{0.559}          & \textbf{0.568} \\
HotpotQA                                              & \multicolumn{1}{c|}{\textbf{0.603}} & 0.58           & \multicolumn{1}{c|}{0.391} & \multicolumn{1}{c|}{0.456} & \multicolumn{1}{c|}{0.584}       & \multicolumn{1}{c|}{0.534} & 0.593          & \multicolumn{1}{c|}{0.535}    & \multicolumn{1}{c|}{0.579}     & \multicolumn{1}{c|}{0.591}          & 0.599          \\
FiQA-2018                                             & \multicolumn{1}{c|}{0.236}          & 0.291          & \multicolumn{1}{c|}{0.112} & \multicolumn{1}{c|}{0.295} & \multicolumn{1}{c|}{0.300}       & \multicolumn{1}{c|}{0.308} & 0.317          & \multicolumn{1}{c|}{0.349}    & \multicolumn{1}{c|}{0.424}     & \multicolumn{1}{c|}{0.444}          & \textbf{0.467} \\
Signal-1M                                             & \multicolumn{1}{c|}{\textbf{0.330}} & 0.307          & \multicolumn{1}{c|}{0.155} & \multicolumn{1}{c|}{0.249} & \multicolumn{1}{c|}{0.289}       & \multicolumn{1}{c|}{0.281} & 0.274          & \multicolumn{1}{c|}{0.261}    & \multicolumn{1}{c|}{0.265}     & \multicolumn{1}{c|}{0.268}          & 0.273          \\
Trec-News                                             & \multicolumn{1}{c|}{0.398}          & \textbf{0.42}  & \multicolumn{1}{c|}{0.161} & \multicolumn{1}{c|}{0.382} & \multicolumn{1}{c|}{0.377}       & \multicolumn{1}{c|}{0.396} & 0.393          & \multicolumn{1}{c|}{0.337}    & \multicolumn{1}{c|}{0.343}     & \multicolumn{1}{c|}{0.350}          & 0.346          \\
Robust04                                              & \multicolumn{1}{c|}{0.408}          & 0.437          & \multicolumn{1}{c|}{0.252} & \multicolumn{1}{c|}{0.392} & \multicolumn{1}{c|}{0.427}       & \multicolumn{1}{c|}{0.362} & 0.391          & \multicolumn{1}{c|}{0.437}    & \multicolumn{1}{c|}{0.470}     & \multicolumn{1}{c|}{0.479}          & \textbf{0.506} \\
ArguAna                                               & \multicolumn{1}{c|}{0.315}          & 0.349          & \multicolumn{1}{c|}{0.175} & \multicolumn{1}{c|}{0.415} & \multicolumn{1}{c|}{0.429}       & \multicolumn{1}{c|}{0.493} & 0.233          & \multicolumn{1}{c|}{0.511}    & \multicolumn{1}{c|}{0.525}     & \multicolumn{1}{c|}{0.531}          & \textbf{0.540} \\
Touché-2020                                           & \multicolumn{1}{c|}{\textbf{0.367}} & 0.347          & \multicolumn{1}{c|}{0.131} & \multicolumn{1}{c|}{0.240} & \multicolumn{1}{c|}{0.162}       & \multicolumn{1}{c|}{0.182} & 0.202          & \multicolumn{1}{c|}{0.205}    & \multicolumn{1}{c|}{0.219}     & \multicolumn{1}{c|}{0.230}          & 0.256          \\
Quora                                                 & \multicolumn{1}{c|}{0.789}          & 0.802          & \multicolumn{1}{c|}{0.248} & \multicolumn{1}{c|}{0.852} & \multicolumn{1}{c|}{0.835}       & \multicolumn{1}{c|}{0.830} & 0.854          & \multicolumn{1}{c|}{0.881}    & \multicolumn{1}{c|}{0.890}     & \multicolumn{1}{c|}{0.890}          & \textbf{0.892} \\
DBPedia-entity                                        & \multicolumn{1}{c|}{0.313}          & 0.331          & \multicolumn{1}{c|}{0.263} & \multicolumn{1}{c|}{0.281} & \multicolumn{1}{c|}{0.384}       & \multicolumn{1}{c|}{0.328} & 0.392          & \multicolumn{1}{c|}{0.347}    & \multicolumn{1}{c|}{0.391}     & \multicolumn{1}{c|}{0.396}          & \textbf{0.408} \\
SCIDOCS                                               & \multicolumn{1}{c|}{0.158}          & \textbf{0.162} & \multicolumn{1}{c|}{0.077} & \multicolumn{1}{c|}{0.122} & \multicolumn{1}{c|}{0.149}       & \multicolumn{1}{c|}{0.143} & 0.145          & \multicolumn{1}{c|}{0.149}    & \multicolumn{1}{c|}{0.158}     & \multicolumn{1}{c|}{0.159}          & 0.161          \\
Fever                                                 & \multicolumn{1}{c|}{0.753}          & 0.714          & \multicolumn{1}{c|}{0.562} & \multicolumn{1}{c|}{0.669} & \multicolumn{1}{c|}{0.700}       & \multicolumn{1}{c|}{0.669} & \textbf{0.771} & \multicolumn{1}{c|}{0.660}    & \multicolumn{1}{c|}{0.712}     & \multicolumn{1}{c|}{0.717}          & 0.740          \\
Climate-Fever                                         & \multicolumn{1}{c|}{0.213}          & 0.201          & \multicolumn{1}{c|}{0.148} & \multicolumn{1}{c|}{0.198} & \multicolumn{1}{c|}{0.228}       & \multicolumn{1}{c|}{0.175} & 0.184          & \multicolumn{1}{c|}{0.241}    & \multicolumn{1}{c|}{0.262}     & \multicolumn{1}{c|}{\textbf{0.270}} & 0.267          \\
SciFact                                               & \multicolumn{1}{c|}{0.665}          & \textbf{0.675} & \multicolumn{1}{c|}{0.318} & \multicolumn{1}{c|}{0.507} & \multicolumn{1}{c|}{0.643}       & \multicolumn{1}{c|}{0.644} & 0.671          & \multicolumn{1}{c|}{0.600}    & \multicolumn{1}{c|}{0.639}     & \multicolumn{1}{c|}{0.635}          & 0.662          \\
CQADupStack                                           & \multicolumn{1}{c|}{0.299}          & 0.325          & \multicolumn{1}{c|}{0.153} & \multicolumn{1}{c|}{0.296} & \multicolumn{1}{c|}{0.314}       & \multicolumn{1}{c|}{0.347} & 0.350          & \multicolumn{1}{c|}{0.357}    & \multicolumn{1}{c|}{0.384}     & \multicolumn{1}{c|}{0.388}          & \textbf{0.399} \\ \midrule
Avg                                                   & \multicolumn{1}{c|}{0.413}          & 0.429          & \multicolumn{1}{c|}{0.234} & \multicolumn{1}{c|}{0.389} & \multicolumn{1}{c|}{0.414}       & \multicolumn{1}{c|}{0.410} & 0.429          & \multicolumn{1}{c|}{0.416}    & \multicolumn{1}{c|}{0.444}     & \multicolumn{1}{c|}{0.452}          & \textbf{0.457} \\
Avg w/o MS Marco                                      & \multicolumn{1}{c|}{0.423}          & {\underline{0.434}}    & \multicolumn{1}{c|}{0.237} & \multicolumn{1}{c|}{0.389} & \multicolumn{1}{c|}{{\underline{0.415}}} & \multicolumn{1}{c|}{0.410} & 0.431          & \multicolumn{1}{c|}{0.416}    & \multicolumn{1}{c|}{0.445}     & \multicolumn{1}{c|}{0.453}          & \textbf{0.458} \\ \bottomrule
\end{tabular}
  }
  \caption{NDCG@10 on the BEIR benchmark. The best result on a given dataset is marked in bold. GTR models are pre-trained on CommunityQA dataset and the complete MS Marco dataset. GTR models achieve better NDCG when increasing size from Base to XXL, outperforming the previous best sparse model DocT5Query and dense retrieval model TAS-B.}
  \label{tab:all_ndcg}
\end{table*}

\subsection{Configurations}
We implement GTR models in JAX\footnote{\url{https://github.com/google/jax}} and train them on Cloud TPU-V8.
We consider different sizes of the T5 transformer~\cite{transformer} architecture including Base, Large, XL and XXL.
Their number of parameters are listed in \cref{tab:param}.

Note that we only use the encoder portion of the T5 models and thus the number of parameters are less than half of the full model size.
We use the off-the-shelf checkpoints as the initial parameters and use the same sentencepiece vocabulary model.\footnote{\url{https://github.com/google-research/text-to-text-transfer-transformer}}

During pre-training and fine-tuning, we set the batch size to 2048 and use a softmax temperature $\tau$ of 0.01. 
We use Adafactor optimizer \citep{shazeer2018adafactor} and set the initial learning rate to 1e-3 with a linear decay.
We train the model for 800K steps and 20K steps for the pre-training and fine-tuning stages, respectively.

For fine-tuning, we use the hard negatives released by RocketQA \citep{qu2021rocketqa} when fine-tuning with MS Marco data and the hard negatives release by \citep{lu2021multi} for Natural Questions, which were proven to lead to better retriever performance. By default, we use the complete MS Marco dataset and the NQ dataset for fine-tuning.

When evaluating on the BEIR benchmark, we use sequences of 64 tokens for the questions and 512 for the documents in all datasets except Trec-News, Robust-04 and ArguAna. In particular, we set the document length to 768 for Trec-News and Robust-04 while setting the question length to 512 for ArguAna, in accordance to the average query and document lengths in these datasets.

\subsection{Models for comparison}

We consider various baselines, including sparse retrieval models: BM25, DocT5Query, and dense retrieval models: DPR, ANCE, TAS-B, and GenQ \citep{thakur2021beir}. 

We conduct experiments on four different sizes of our GTR models (GTR-Base, GTR-Large, GTR-XL, and GTR-XXL).
We consider three different settings for GTR to investigate the scaling up effect for different training stages:
\begin{itemize}
    \item GTR. This is the full GTR model that conducts both pre-training and fine-tuning.
    \item GTR-FT. This is a fine-tune only version of GTR where the T5 dual encoders on the MS Marco dataset.
    \item GTR-PT. This is a pre-training only version of GTR where the T5 dual encoders is only pre-trained on the CommunityQA dataset.

\end{itemize}

We evaluate baseline and our models on the BEIR generalization task~\cite{thakur2021beir} as discussed in section~\ref{sec:beir},. We consider two main retrieval metrics: NDCG@10 and Recall@100 following BEIR~\cite{thakur2021beir}. Due to space limitations, we report NDCG@10 in the main section of the paper and include Recall@100 results in  \cref{sec:appendix}. 

\section{Evaluation Results}

We present three groups of experiments to study the a) in-domain performance of the GTR models on MS Marco, b) their out-of-domain generalization performance on BEIR, and c) their data efficiency. 
\subsection{Results on MS Marco}
We first analyze in-domain performance based on the evaluation results on MS Marco.
As show in \cref{tab:all_ndcg}, with scaling up, the models achieve consistent improvement on NDCG@10. We observe similar improvements on other evaluation metrics including MRR@10 and Recall@1000 and reported the numbers in  \cref{tab:msmarco} of \cref{sec:appendix}. 
This shows that increasing model capacity leads to better in-domain performance.

\begin{figure}[t]
  \centering
  \includegraphics[width=\linewidth]{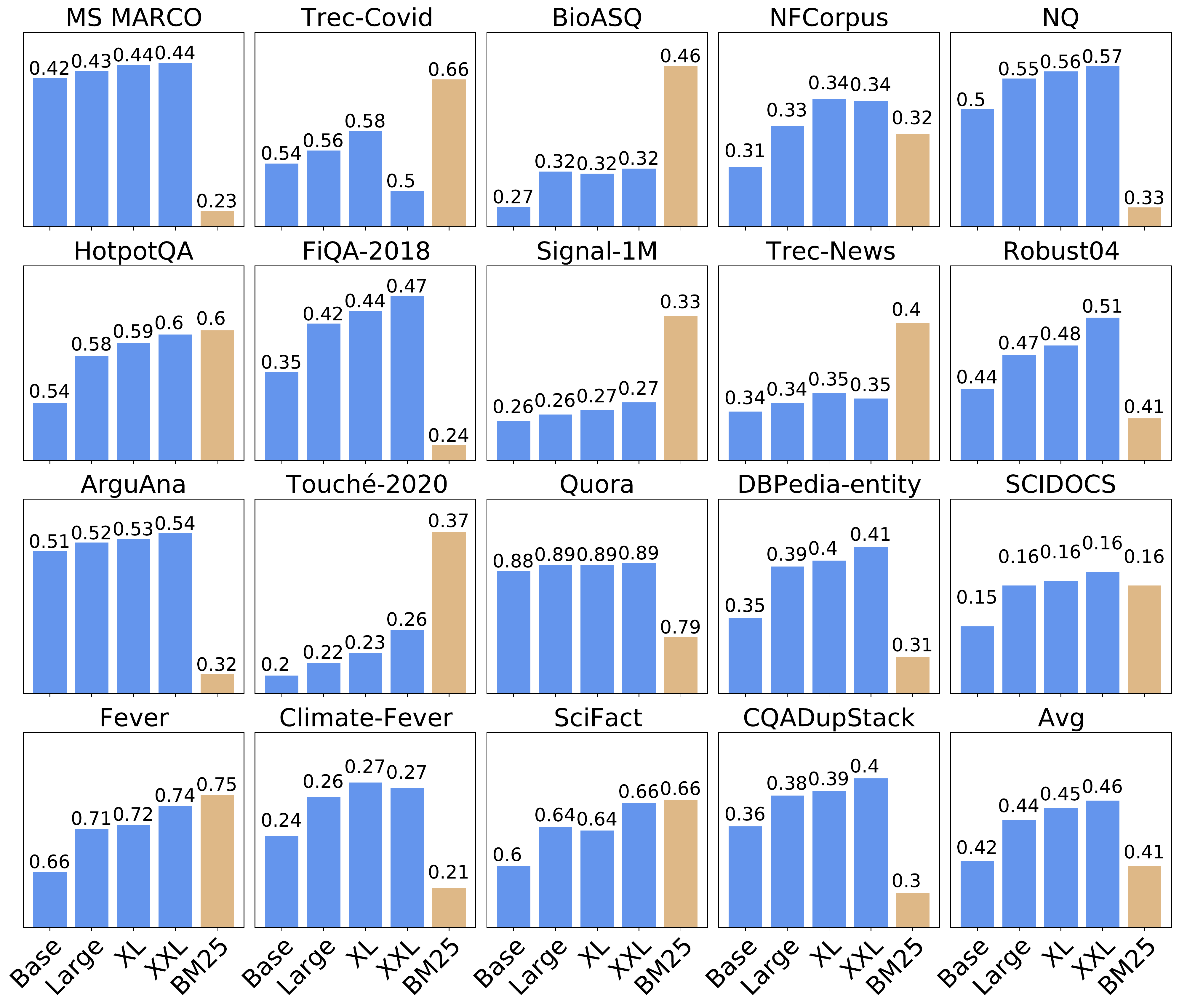}
  \caption{Comparison with BM25 on NDCG@10. The GTR-Base model outperforms BM25 on 9 datasets and the larger GTR models continue to improve on these 9 tasks. The GTR-XXL model catches up or surpasses BM25 on the other 5 datasets and only under-performs on 5 of the remaining tasks.}
  \label{fig:comparison-bm25}
\end{figure}

\subsection{Results on BEIR generalization tasks}

The next set of experiments investigates the effect of increasing model capacity on out-of-domain (OOD) performance. 

As shown in \cref{tab:all_ndcg}, we observe a clear gain on out-of-domain performance in terms of NDCG@10 when the model size increases. The GTR-Large model already outperforms the previous best dense retrieval model TAS-B as well as the best sparse model DocT5Query.  Scaling up to GTR-XXL leads to another jump in retrieval performance. Similar improvements are found on Recall@100 as shown in the Appendix's \cref{tab:all_recall}. On average, the scaling up process demonstrates an encouraging ascending trend that eventually outperforms all baseline methods on all evaluation metrics. This confirms that scaling up is a valid path towards generalizability. 

Previously, dual encoders failed to match the performance of BM25 for tasks that require better lexical matching capabilities.
Thus, we wanted to investigate what kind of tasks can get improved by scaling up the model size. \Cref{fig:comparison-bm25} presents a detailed comparison of all sizes of GTR models against the BM25 baseline.

For tasks like NQ where dual encoders have been previously shown to be more effective than BM25, increasing the model size continues to advance the performance of dual encoders. This suggests scaling up can further boost the head start of dense models over sparse models on these datasets.

For tasks like BioASQ and NFCorpus, where dual encoders previously struggled to match the performance of BM25 for inherent reasons, we discovered that scaling up consistently improves the retrieval performance. In particular, for NFCorpus, our Base model under-performs BM25 but the XL model outperforms BM25 by 5.5\% (0.343 vs. 0.325). 
This exciting finding verifies our assumption that scaling up can further exploit the powerful semantic matching capabilities of the dual encoder models and enable them to ultimately outperform BM25.

\begin{table}[t]
    \small
    \centering
    \resizebox{\linewidth}{!}{%
    \begin{tabular}{l|rr|rrr}
    \toprule
      & \multicolumn{2}{c}{GTR-FT}  & \multicolumn{3}{c}{GTR} \\
      \midrule
    Ratio of data& Large & XL & Large & XL & XXL \\
    \midrule
    \multicolumn{6}{c}{NDCG@10 on MS Marco} \\
    \midrule
    10\% & 0.402 & 0.397  & 0.428  & 0.426 & - \\
    100\% & \underline{0.415} & \underline{0.418} & \underline{0.430} & \underline{0.439} & \underline{0.430} \\
    \midrule
    \multicolumn{6}{c}{Zero-shot average NDCG@10 w/o MS Marco} \\ \midrule
    10\% & \textbf{0.413} & 0.418 & \textbf{0.452} & \textbf{0.462} & \textbf{0.465} \\
    100\% & 0.412 & \textbf{0.433} & 0.445 & 0.453 & 0.458 \\
    \bottomrule
    \end{tabular}
    }
    \caption{Comparisons of NDCG@10 for GTR models trained with different amount of fine-tuning data. With only 10\% of the MS Marco data, both GTR-FT and GTR large and XL models achieve slightly worse in-domain performance; meanwhile they obtain comparable or even superior out-of-domain performance than using the complete MS Marco data.}
  \label{tab:data_negatives}
\end{table}

\subsection{Data efficiency for large retrievers}

To better understand the data efficiency for large dual encoders, we trained models using different portions of the MS Marco dataset during fine-tuning. In particular, we sampled a subset of the training data by keeping only 10\% of the training queries as well as their relevant (positive) passages and irrelevant (hard negative) passages.

As shown in \cref{tab:data_negatives}, using 10\% of training data reduces the in-domain performance of the  GTR models on MS Marco. For the GTR-FT (fine-tuning only) models, using 10\% of the data leads to a mixed result of out-of-domain performance.

On the other hand, for full GTR models, using 10\% of the MS Marco dataset is sufficient for fine-tuning. 
In particular, the GTR-Large, XL and XXL models achieve comparable or even better OOD performance than fine-tuning on the complete MS Marco dataset. This might suggest that GTR models have the benefit of data efficiency and could use less training data for domain adaptation.

\section{Ablation Study and Analysis}
In this section we present ablations and analysis to further understand the effects of scaling up, the impact of fine-tuning and pre-training, and %
the trends of the GTR model on different experimental conditions.
\subsection{Effect of scaling up for different training stages}

\begin{table}[t]
  \small
  \centering
\begin{tabularx}{0.9\linewidth}{l|r r r}
\toprule
        & GTR-FT & GTR-PT & GTR \\
        \midrule
Fine-tuning & \cmark  & \xmark & \cmark \\
\midrule
 \multicolumn{4}{c}{NDCG@10 on MS Marco} \\
\midrule
Base    & 0.400 & 0.258   & 0.420 \\
Large   & 0.415 & \underline{0.262}   & 0.430 \\
XL      & 0.418 & 0.259   & 0.439 \\
XXL     & \underline{0.422} & 0.252   & \underline{0.442} \\
\midrule
 \multicolumn{4}{c}{Zero-shot average NDCG@10 w/o MS Marco} \\
\midrule
Base    & 0.387 & 0.295   & 0.416 \\
Large   & 0.412 & 0.315   & 0.445 \\
XL      & \textbf{0.433} & 0.315   & 0.453 \\
XXL     & 0.430 & \textbf{0.332}   & \textbf{0.458} \\
\bottomrule
  \end{tabularx}
  \caption{Comparisons (NDCG@10) of the models trained with and without pre-training and fine-tuning. Notably, the GTR-FT XL model already achieves an average zero-shot NDCG@10 of 0.433, which outperforms the previous best dual encoder model TAS-B (NDCG@10=0.415).}
  \label{tab:comparisons}
\end{table}

The first ablation study aims to investigate how scaling up effects dual encoder pre-training and fine-tuning. Results are listed in \cref{tab:comparisons}.

For fine-tuning only models, scaling up benefits both in-domain and out-of-domain performance.
For pre-training only models, the improvement on in-domain performance is not obvious; meanwhile for out-of-domain tasks, scaling up also improves the generalization.
Finally with both pre-training and fine-tuning, GTR models consistently improve over GTR-FT models of all sizes. This shows the power of combining scaling up and a generic pre-training stage.

\subsection{Importance of the fine-tuning dataset}
In \cref{tab:comparisons}, we compare GTR and GTR-PT on the BEIR benchmark to understand the importance of fine-tuning on MS Marco. The table shows that there is a clear gap between GTR models before and after fine-tuning. The result shows the necessity of leveraging a high quality dataset (e.g.~search data) to fine-tune the dual encoders.

In \cref{tab:nq}, we compare fine-tuning GTR on NQ instead of MS Marco. Compared to MS Marco, NQ only covers Wikipedia documents and is much smaller in size, which allows us to investigate the performance of GTR when fine-tuned on a less generalizable dataset. In addition, fine-tuning on NQ can give us a fair comparison with DPR. 

As shown in \cref{tab:nq}, the GTR-base model fine-tuned on NQ outperforms the original DPR model, which uses a BERT-Base model as the encoder backbone. This demonstrates the effectiveness of our pre-training on the Web dataset as well as the hard negatives introduced from \citet{lu2021multi} for NQ.  Fine-tuning on NQ leads to inferior performance compared to fine-tuning on MS Marco, which is consistent with prior work~\cite{thakur2021beir}. However, importantly, scaling up GTR size improves zero-shot performance on BEIR when fine-tuning on NQ.
This shows that the benefit of scaling up holds for different fine-tuning datasets. Furthermore, when scaling from Large to XL, we observe a larger gain when fine-tuning with NQ than with MS Marco, indicating that scaling up helps more when using weaker fine-tuning data.

\begin{table}[]
    \small
    \centering
    \begin{tabularx}{0.95\linewidth}{l|l|X}
    \toprule
    Model & Fine-tuning dataset & Zero-shot average NDCG@10 \\
    \midrule
    DPR & NQ & 0.237 \\
    GTR-Base & NQ  & 0.360 \\
    GTR-Large & NQ  & 0.379 \\
    GTR-XL & NQ &  \textbf{0.407} \\
    \midrule
    GTR-Large & MS Marco & 0.445 \\
    GTR-XL & MS Marco & \underline{0.453} \\
    \bottomrule
    \end{tabularx}
    \caption{Comparisons of GTR models fine-tuned on MS Marco and NQ. We report the zero-shot average NDCG@10. Scaling up improves model performance both on NQ and MS Marco.}
    \label{tab:nq}
\end{table}

\subsection{Document length vs model capacity}

\begin{figure}[t]
  \centering
  \includegraphics[width=\linewidth]{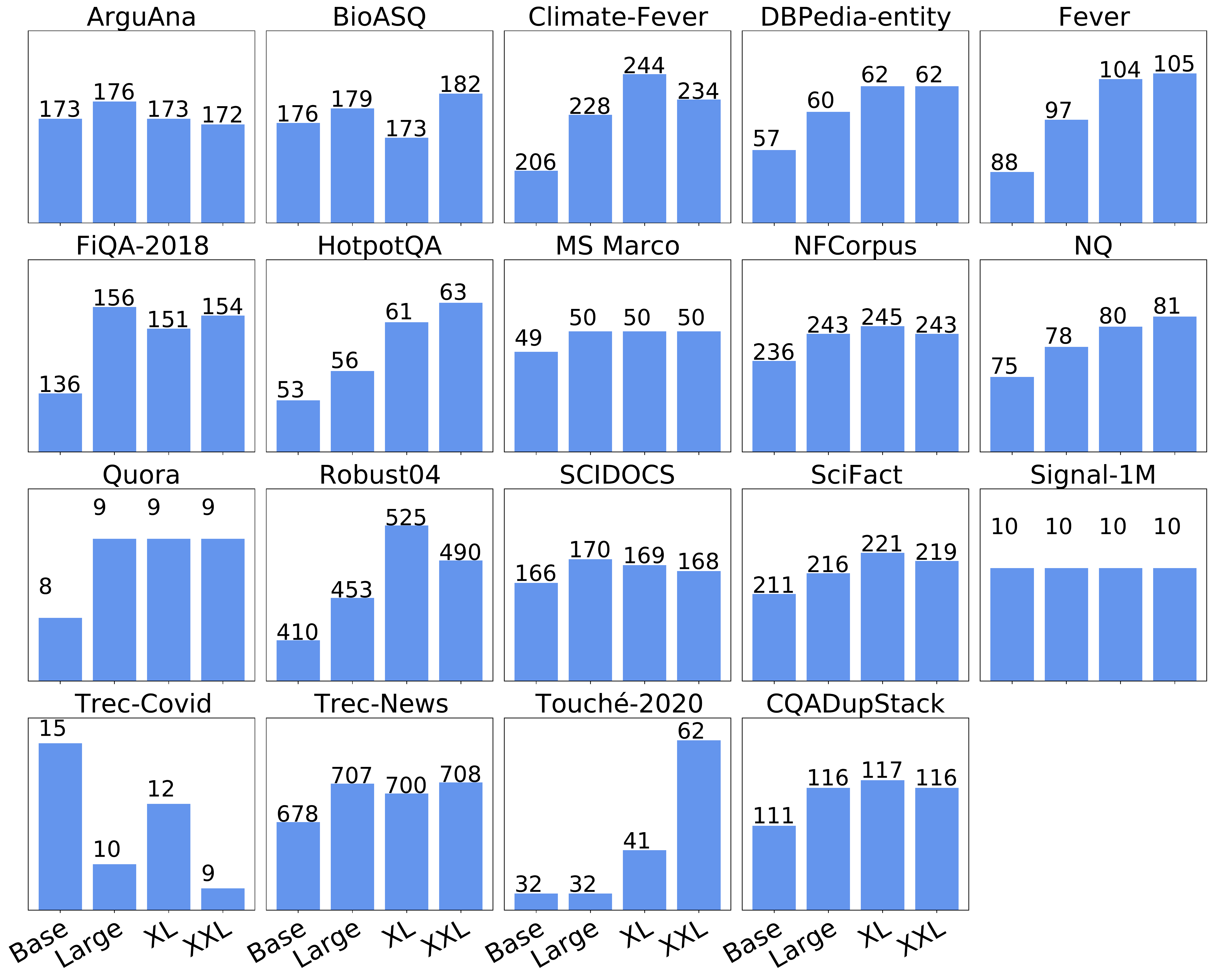}
  \caption{Median lengths (in words) of top-10 retrieved documents for all queries.
  }
  \label{fig:doc-lengths}
\end{figure}

Previously, BEIR has shown that models trained with cosine similarity prefer short documents while those trained with dot-product prefer long documents~\cite{thakur2021beir}. We investigate whether scaling up affect this observation. Specifically, we compute the median lengths (in words) of the top-10 retrieved documents for all queries. Results are shown in \cref{fig:doc-lengths}.

Though all GTR models are trained using cosine similarity, we found that scaling up the model size has influence over the lengths of retrieved documents.
We observe an increasing trend of document length for DB-Pedia, Fever, HotpotQA, Signal-1M, Trec-News, and Web-Touche2020 with scaling up. In particular, for Web-Touche2020, the lengths of the retrieved documents grow drastically as the models scale up: The largest GTR-XXL retrieves documents that are on average twice as long compared with the smallest GTR-Base. This plays in our favor since \citet{thakur2021beir} show that the majority of relevant documents in Web-Touche2020 are longer.

On the other hand, the only exception we observe is the Trec-Covid dataset, where GTR-XXL model retrieves much shorter documents than those retrieved by the smaller size counterparts. This may explain the inferior performance of GTR-XXL on Trec-Covid shown in \cref{tab:all_ndcg} and \cref{tab:all_recall}. We leave it as future work to explore the effects of using the dot-product as similarity function for large dual encoders.

\section{Related Work}

\paragraph{Neural information retrieval.}
Document retrieval is an important task in the NLP and information retrieval (IR) communities. The goal  is to find the relevant document from a large corpus given a query.
Traditionally, lexical based approaches trying to match the query and document based on term overlap, such as TF-IDF and BM25~\cite{bm25}, have achieved great success in this task.
Recently, neural based approaches, which go beyond the simple term matching, are being quickly adopted by the community and achieve state-of-the-art performance on multiple retrieval tasks, such as passage retrieval~\cite{karpukhin-etal-2020-dense}, question answering~\cite{ahmad-etal-2019-reqa}, conversational question answering~\cite{Qu2020OpenRetrievalCQ} and bitext retrieval~\cite{feng2020languageagnostic}.

\paragraph{Dual encoders for neural retrieval.}
Dual encoders have demonstrated to be one type of neural retrievers that can achieve great performance compared to traditional sparse models such as BM25 for a wide range of retrieval tasks \citep{karpukhin-etal-2020-dense,Gillick2018EndtoEndRI}. 
One key aspect to their success is the adoption of pre-trained language models, which enables the dual encoders to have backbone contextual embeddings to initialize from. Other techniques such as negative mining \citep{xiong2020approximate,lu2021multi,sachan-etal-2021-end} and large training batch sizes \citep{qu2021rocketqa} have also shown great effectiveness. However, few of the previous works have discussed the effect of the backbone model's capacity. 

\paragraph{Zero-shot neural retrieval.}
Recent works have shown great improvement under the zero-shot setting for dual encoders by leveraging distillation and synthetic data generation \citep{thakur2021beir,hofstatter2021efficiently,ma2020zero}.
Both of these techniques, and scaling up backbone models, are effective ways to close the gap between dual encoders and the upper bound of the single-product approaches with fixed-dimension embeddings.
On the other hand, multi-vector approaches introduce more interactions between dense embeddings, which could also benefit from scaling up the backbone multi-vector encoders.  
We hope that our observation about scaling up model sizes for single dot-product based methods can be combined with these techniques and further push the frontier of neural retrieval models.

\section{Inference latency}

One caveat for scaling up model size is the increment in the latency overhead. We investigate the inference speed in terms of microseconds (ms) for all GTR models with batch size 1 and input length 128. We found the latency increases from 17 ms, 34 ms, 96 ms to 349 ms. 
The GTR-Base model has close latency compared to TAS-B while the largest GTR-XXL model has a similar latency to the re-ranking models \citep{thakur2021beir}.
With the recent work towards making large models efficient from angles such as sparsity, distillation and prompt-tuning, we hope the inference time for large dual encoders can be significantly reduced in the future.

\section{Conclusion}

This paper presents the Generalizable T5 Retriever (GTR), a scaled-up dual encoder model with a fixed-size bottleneck layer.
We show that scaling up the model size brings significant improvement on retrieval performance across the board on the BEIR zero-shot retrieval benchmark, especially for out-of-domain generalization. 
The GTR-XXL model achieves state-of-the-art performance on the BEIR benchmark, outperforming many models that use earlier interactions between queries and documents.
This sheds light on the research direction to keep improving the single vector representation model through better backbone encoders. 
The findings here are also complementary with other recent works that improve the dual encoder training, including distilling from a ranker~/~scorer model, using a better contrasting pre-training objective and scaling up the encoders for multi-vector retrieval models.

\section*{Acknowledgments}
We thank Chris Tar and Don Metzler for feedback and suggestions.

\bibliography{custom}
\bibliographystyle{acl_natbib}

\appendix

\clearpage

\section{More results}
\label{sec:appendix}

\subsection{Comparisons on MS Marco}

\Cref{tab:msmarco} shows the comparisons of GTR models and the baselines. Note that the best RocketQA model used additional augmented data other than MS Marco to improve the model performance while all others do not. Our best GTR-XXL models outperforms RocketQA on both MRR and recall.

\begin{table}[]
  \small
  \centering
  \resizebox{\linewidth}{!}{%
\begin{tabular}{l|r|r|r}
\toprule
Model & NDCG@10 & MRR@10 & Recall@1000 \\
\midrule
ANCE  & 0.388 & 0.330 & 0.959      \\
TAS-Balanced & 0.408 & 0.340 & 0.975      \\
ColBERT      & 0.401 & 0.360 & 0.968      \\
RocketQA     & / & 0.370 & 0.979      \\
GTR-Base     & 0.420 & 0.366 &  0.983  \\
GTR-Large    & 0.430 & 0.379 & \textbf{0.991} \\
GTR-XL       & 0.439 & 0.385 & 0.989 \\
GTR-XXL      & \textbf{0.442} & \textbf{0.388} & 0.990 \\
\bottomrule
  \end{tabular}
 }
  \caption{Comparisons of different models on MS Marco. Scaling up can improve GTR models' in-domain performance.}
  \label{tab:msmarco}
\end{table}

\subsection{Comparison of different dual encoder pre-training strategies}

In a concurrent work~\citep{anonymous2022contrastive}, researchers  proposed to conduct contrastive learning (CL) pre-training for improving the generalizability of neural retrievers.
The paired data for contrastive training is constructed from C4 and Wiki dataset in an unsupervised way.
In particular, they construct pairs by randomly choosing two spans from a single document and conduct word deletion or replacement to each span.
We compare the performance of our GTR models to their models to gain insights into different pretraining strategies for dual encoders.

\begin{figure}[]
  \centering
  \includegraphics[width=\linewidth]{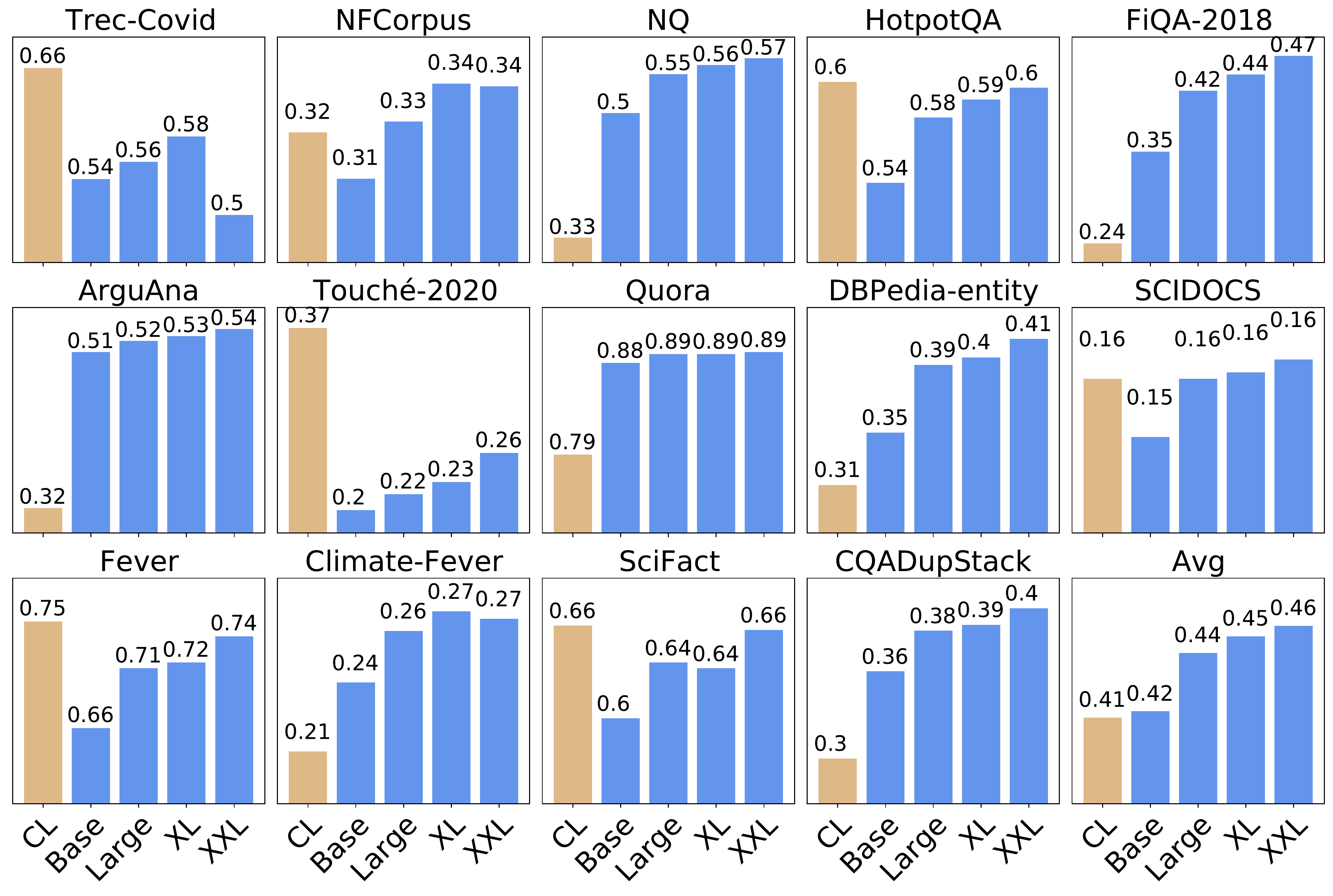}
  \caption{Comparison with \citet{anonymous2022contrastive} on NDCG@10. ``CL'' denotes \citet{anonymous2022contrastive} with contrastive learning on C4 and Wiki while others denote our GTR models with different sizes. Note that they only report results on 15 datasets of the BEIR benchmark.}
  \label{fig:comparison-cl-pretrain}
\end{figure}

As shown in Figure~\ref{fig:comparison-cl-pretrain}, on over half of the datasets, models with our pre-training approach under-perform CL-Pretrain with the base size; while as the model size increases, GTR-Large and -XXL models show significant gains over CL-Pretrain. The best GTR-XXL model achieves 0.49 for NDCG@10 on average while CL-Pretrain achieves 0.46.
This demonstrates that scaling up can mitigate the disadvantage of the potentially inferior pre-training approach.
Note that our pre-training is additive to CL-Pretrain and we can leverage the pre-training on C4 and Wiki to further improve the results. We leave this exploration as future work.

\subsection{Recall on BEIR}

\Cref{tab:all_recall} presents the Recall@100 of GTR models and the baselines. Similar to NDCG@10, we observe that scaling up dual encoders lead to significant gains on the BEIR benchmark in terms of recall.

\begin{table*}[h]
  \small
  \centering
  \resizebox{\linewidth}{!}{%
\begin{tabular}{@{}l|cc|ccccc|cccc@{}}
\toprule
\multicolumn{1}{c|}{\multirow{2}{*}{Recall@10 / Model}} & \multicolumn{2}{c|}{Lexical / Sparse}                   & \multicolumn{5}{c|}{Dense}                                                                                                              & \multicolumn{4}{c}{Ours}                                                                                              \\ \cmidrule(l){2-12} 
\multicolumn{1}{c|}{}                                   & \multicolumn{1}{c|}{BM25}           & docT5query        & \multicolumn{1}{c|}{DPR}   & \multicolumn{1}{c|}{ANCE}  & \multicolumn{1}{c|}{TAS-B}             & \multicolumn{1}{c|}{GenQ}  & ColBERT & \multicolumn{1}{c|}{GTR-Base} & \multicolumn{1}{c|}{GTR-Large} & \multicolumn{1}{c|}{GTR-XL}         & GTR-XXL        \\ \midrule
MS Marco                                                & \multicolumn{1}{c|}{0.658}          & 0.819             & \multicolumn{1}{c|}{0.552} & \multicolumn{1}{c|}{0.852} & \multicolumn{1}{c|}{0.884}             & \multicolumn{1}{c|}{0.884} & 0.865   & \multicolumn{1}{c|}{0.898}    & \multicolumn{1}{c|}{0.908}     & \multicolumn{1}{c|}{0.911}          & \textbf{0.916} \\
Trec-Covid                                              & \multicolumn{1}{c|}{0.498}          & \textbf{0.541}    & \multicolumn{1}{c|}{0.212} & \multicolumn{1}{c|}{0.457} & \multicolumn{1}{c|}{0.387}             & \multicolumn{1}{c|}{0.456} & 0.464   & \multicolumn{1}{c|}{0.411}    & \multicolumn{1}{c|}{0.434}     & \multicolumn{1}{c|}{0.457}          & 0.407          \\
BioASQ                                                  & \multicolumn{1}{c|}{\textbf{0.714}} & 0.646             & \multicolumn{1}{c|}{0.256} & \multicolumn{1}{c|}{0.463} & \multicolumn{1}{c|}{0.579}             & \multicolumn{1}{c|}{0.627} & 0.645   & \multicolumn{1}{c|}{0.441}    & \multicolumn{1}{c|}{0.490}     & \multicolumn{1}{c|}{0.483}          & 0.483          \\
NFCorpus                                                & \multicolumn{1}{c|}{0.250}          & 0.253             & \multicolumn{1}{c|}{0.208} & \multicolumn{1}{c|}{0.232} & \multicolumn{1}{c|}{0.280}             & \multicolumn{1}{c|}{0.280} & 0.254   & \multicolumn{1}{c|}{0.275}    & \multicolumn{1}{c|}{0.298}     & \multicolumn{1}{c|}{\textbf{0.318}} & 0.300          \\
NQ                                                      & \multicolumn{1}{c|}{0.760}          & 0.832             & \multicolumn{1}{c|}{0.880} & \multicolumn{1}{c|}{0.836} & \multicolumn{1}{c|}{0.903}             & \multicolumn{1}{c|}{0.862} & 0.912   & \multicolumn{1}{c|}{0.893}    & \multicolumn{1}{c|}{0.930}     & \multicolumn{1}{c|}{0.936}          & \textbf{0.946} \\
HotpotQA                                                & \multicolumn{1}{c|}{0.740}          & 0.709             & \multicolumn{1}{c|}{0.591} & \multicolumn{1}{c|}{0.578} & \multicolumn{1}{c|}{0.728}             & \multicolumn{1}{c|}{0.673} & 0.748   & \multicolumn{1}{c|}{0.676}    & \multicolumn{1}{c|}{0.725}     & \multicolumn{1}{c|}{0.739}          & \textbf{0.752} \\
FiQA-2018                                               & \multicolumn{1}{c|}{0.539}          & 0.598             & \multicolumn{1}{c|}{0.342} & \multicolumn{1}{c|}{0.581} & \multicolumn{1}{c|}{0.593}             & \multicolumn{1}{c|}{0.618} & 0.603   & \multicolumn{1}{c|}{0.670}    & \multicolumn{1}{c|}{0.742}     & \multicolumn{1}{c|}{0.755}          & \textbf{0.780} \\
Signal-1M                                               & \multicolumn{1}{c|}{\textbf{0.370}} & 0.351             & \multicolumn{1}{c|}{0.162} & \multicolumn{1}{c|}{0.239} & \multicolumn{1}{c|}{0.304}             & \multicolumn{1}{c|}{0.281} & 0.283   & \multicolumn{1}{c|}{0.263}    & \multicolumn{1}{c|}{0.261}     & \multicolumn{1}{c|}{0.268}          & 0.268          \\
Trec-News                                               & \multicolumn{1}{c|}{0.422}          & 0.439             & \multicolumn{1}{c|}{0.215} & \multicolumn{1}{c|}{0.398} & \multicolumn{1}{c|}{0.418}             & \multicolumn{1}{c|}{0.412} & 0.367   & \multicolumn{1}{c|}{0.475}    & \multicolumn{1}{c|}{0.525}     & \multicolumn{1}{c|}{0.512}          & \textbf{0.544} \\
Robust04                                                & \multicolumn{1}{c|}{\textbf{0.375}} & 0.357             & \multicolumn{1}{c|}{0.211} & \multicolumn{1}{c|}{0.274} & \multicolumn{1}{c|}{0.331}             & \multicolumn{1}{c|}{0.298} & 0.31    & \multicolumn{1}{c|}{0.324}    & \multicolumn{1}{c|}{0.365}     & \multicolumn{1}{c|}{0.364}          & 0.372          \\
ArguAna                                                 & \multicolumn{1}{c|}{0.942}          & 0.972             & \multicolumn{1}{c|}{0.751} & \multicolumn{1}{c|}{0.937} & \multicolumn{1}{c|}{0.942}             & \multicolumn{1}{c|}{0.978} & 0.914   & \multicolumn{1}{c|}{0.974}    & \multicolumn{1}{c|}{0.978}     & \multicolumn{1}{c|}{0.980}          & \textbf{0.983} \\
Touché-2020                                             & \multicolumn{1}{c|}{0.538}          & \textbf{0.557}    & \multicolumn{1}{c|}{0.301} & \multicolumn{1}{c|}{0.458} & \multicolumn{1}{c|}{0.431}             & \multicolumn{1}{c|}{0.451} & 0.439   & \multicolumn{1}{c|}{0.281}    & \multicolumn{1}{c|}{0.282}     & \multicolumn{1}{c|}{0.297}          & 0.301          \\
Quora                                                   & \multicolumn{1}{c|}{0.973}          & 0.982             & \multicolumn{1}{c|}{0.470} & \multicolumn{1}{c|}{0.987} & \multicolumn{1}{c|}{0.986}             & \multicolumn{1}{c|}{0.988} & 0.989   & \multicolumn{1}{c|}{0.996}    & \multicolumn{1}{c|}{0.996}     & \multicolumn{1}{c|}{\textbf{0.997}} & \textbf{0.997} \\
DBPedia-entity                                          & \multicolumn{1}{c|}{0.398}          & 0.365             & \multicolumn{1}{c|}{0.349} & \multicolumn{1}{c|}{0.319} & \multicolumn{1}{c|}{\textbf{0.499}}    & \multicolumn{1}{c|}{0.431} & 0.461   & \multicolumn{1}{c|}{0.418}    & \multicolumn{1}{c|}{0.480}     & \multicolumn{1}{c|}{0.480}          & 0.494          \\
SCIDOCS                                                 & \multicolumn{1}{c|}{0.356}          & 0.360             & \multicolumn{1}{c|}{0.219} & \multicolumn{1}{c|}{0.269} & \multicolumn{1}{c|}{0.335}             & \multicolumn{1}{c|}{0.332} & 0.344   & \multicolumn{1}{c|}{0.340}    & \multicolumn{1}{c|}{0.358}     & \multicolumn{1}{c|}{0.358}          & \textbf{0.366} \\
Fever                                                   & \multicolumn{1}{c|}{0.931}          & 0.916             & \multicolumn{1}{c|}{0.840} & \multicolumn{1}{c|}{0.900} & \multicolumn{1}{c|}{0.937}             & \multicolumn{1}{c|}{0.928} & 0.934   & \multicolumn{1}{c|}{0.923}    & \multicolumn{1}{c|}{0.941}     & \multicolumn{1}{c|}{0.944}          & \textbf{0.947} \\
Climate-Fever                                           & \multicolumn{1}{c|}{0.436}          & 0.427             & \multicolumn{1}{c|}{0.390} & \multicolumn{1}{c|}{0.445} & \multicolumn{1}{c|}{0.534}             & \multicolumn{1}{c|}{0.450} & 0.444   & \multicolumn{1}{c|}{0.522}    & \multicolumn{1}{c|}{0.552}     & \multicolumn{1}{c|}{\textbf{0.569}} & 0.556          \\
SciFact                                                 & \multicolumn{1}{c|}{0.908}          & \textbf{0.914}    & \multicolumn{1}{c|}{0.727} & \multicolumn{1}{c|}{0.816} & \multicolumn{1}{c|}{0.891}             & \multicolumn{1}{c|}{0.893} & 0.878   & \multicolumn{1}{c|}{0.872}    & \multicolumn{1}{c|}{0.899}     & \multicolumn{1}{c|}{0.911}          & 0.900          \\
CQADupStack                                             & \multicolumn{1}{c|}{0.606}          & 0.638             & \multicolumn{1}{c|}{0.403} & \multicolumn{1}{c|}{0.579} & \multicolumn{1}{c|}{0.622}             & \multicolumn{1}{c|}{0.654} & 0.624   & \multicolumn{1}{c|}{0.681}    & \multicolumn{1}{c|}{0.714}     & \multicolumn{1}{c|}{0.729}          & \textbf{0.740} \\ \midrule
Avg                                                     & \multicolumn{1}{c|}{0.601}          & 0.615             & \multicolumn{1}{c|}{0.425} & \multicolumn{1}{c|}{0.559} & \multicolumn{1}{c|}{0.610}             & \multicolumn{1}{c|}{0.605} & 0.604   & \multicolumn{1}{c|}{0.596}    & \multicolumn{1}{c|}{0.625}     & \multicolumn{1}{c|}{0.632}          & \textbf{0.634} \\
Avg w/o MS Marco                                        & \multicolumn{1}{c|}{0.598}          & \underline{0.603} & \multicolumn{1}{c|}{0.418} & \multicolumn{1}{c|}{0.543} & \multicolumn{1}{c|}{\underline{0.594}} & \multicolumn{1}{c|}{0.590} & 0.590   & \multicolumn{1}{c|}{0.580}    & \multicolumn{1}{c|}{0.609}     & \multicolumn{1}{c|}{0.616}          & \textbf{0.619} \\ \bottomrule
\end{tabular}
  }
  \caption{Recall@100 on the BEIR benchmark. The best result on a given dataset is marked in bold.}
  \label{tab:all_recall}
\end{table*}

\end{document}